# Identifying the optimal parameters for sprayed and inhaled drug particulates for intranasal targeting of SARS-CoV-2 infection sites

Yueying Lao[1], Diane Joseph-McCarthy[1], Arijit Chakravarty[2], Pallavi A. Balivada[1], Phoebe Ato[1], Nogaye K. Ka[1], Saikat Basu[3] *

[1] *Department of Biomedical Engineering, Boston University, Boston, MA*
[2] *Fractal Therapeutics, Cambridge, MA*
[3] *Department of Mechanical Engineering, South Dakota State University, Brookings, SD*

**Abstract**

Efficacy for COVID-19 treatments can be enhanced significantly through targeting the nasopharynx, which has been shown to be the dominant preliminary infection site for SARS-CoV-2. Although intranasal drugs can be administered easily through drops or sprays, it is difficult to test whether current protocols will deliver the right amount of the drug to this location consistently. We are interested in developing an *in silico* prototyping tool to rapidly identify optimal parameters for intranasal delivery. In this study, we have applied computational fluid dynamics to simulate fluid flow through the nasal cavity and examined particle deposition for a drug formulation, mimicking different delivery methods. The nasal geometry models were derived using digitized and meshed computed tomography (CT) scans of human patients. Using the nasal geometries, we simulated two different airflows: a laminar model at 15 LPM (Liters/min) that simulated resting breathing rate and a Large Eddy Simulation (LES) model used to achieve a higher flow rate of 30 LPM. We were able to run particle tracking simulations for these two airflow schemes to test different drug properties such as particle size. The different injection methods used include surface injection which best replicates an inhaler-based release of particle droplets into the nostril and the cone injection method which best replicates a spray into the nostril. The results of the study suggest that the most optimal drug particle size for targeting the intranasal infection sites is around 6-14 µm.

* Corresponding author; Email: Saikat.Basu@sdstate.edu; saikat25@vt.edu

## 1. Introduction

There is an urgent and immediate need for an alternate treatment of the SARS-CoV-2 infection. This deadly virus has claimed the lives of more than one million people worldwide[1,2]. For infected patients that are symptomatic, the conditions are often unbearable and difficult to live with. Researchers have taken unprecedented steps to quickly find a safe and effective treatment[3]. Even with the current limited scale of vaccines in development, it should be noted that a wide range of approaches are being employed such as inactivated virus, protein subunits, live attenuated virus and gene-editing technology, each with its associated risks of vaccine failure[4]. Additionally, researchers working on vaccines typically go through the tasking process of making the vaccine, testing it in animals, going through the three phases of clinical trials to establish it, completing safety and efficacy testing and finally regulatory approval and registration of vaccine in countries where it would be used. The uncertainty surrounding high-risk efforts such as vaccine development and IV-administered antivirals points to a critical need for an effective and easily deployable therapeutic or prophylaxis agent[5].

Since studies have found that the viral load in the nasopharynx peaks on or before the onset of symptoms, an early intervention method targeting the nasopharynx is imperative for limiting asymptomatic transmission of the disease as well as preventing progression of the disease towards severe illness[6,7]. A second concern is that the mutation rate of SARS-CoV-2 and the nature of the fitness landscape makes the virus comfortable with evolving, potentially resulting in more virulent strains[8]. In this context, exposure to a drug can act as a selection mechanism for resistant strains, reducing the efficacy of a previously successful treatment[9]. A nasal spray would address this concern by targeting the virus in the early stages of the disease and by reducing the risk of mutation within the host. A device localization as well as the drug combination precludes the odds of the virus symptoms getting a foothold on the infected person.

Our proposed computational prototyping tool will assist in identifying optimal parameters of the drug formulation and the delivery device. Developing the drug into a final product will require significant time-cost due to testing delivery methods. However, computational screening significantly reduces the burden of experimental testing and validation, allowing faster development.

## 2. Methods

### Anatomic Nasal Cavity Reconstruction

**Medical Imaging**

The nasal cavity models used in this study were reconstructed from the de-identified computed tomography (CT) scans of two healthy test subjects. Subject 1 is a 37 year-old female (UNC003_Basu) and subject 2 is a 61 year-old female (UNC071_Basu)[10]. The use of these medical scans for computational research was authorized by the Institutional Review Board at University of North Carolina at Chapel Hill.

**Development of *in Silico* Models and Meshing Procedure**

Digital models of nasal airways were constructed from the medical scans using a delineation range of -1024 to -300 Hounsfield units. Additionally, pixels were manually adjusted to improve anatomic accuracy of the model. These initial processing steps were performed using Mimics 18.0 (Materialise, Plymouth, Michigan). The cavities were then meshed into small volume elements using ICEM-CFD 15.0 (ANSYS Inc., Canonsburg, Pennsylvania). To ensure grid-independent solutions, mesh-refinement protocols were used such that each computational grid contained more than 4 million unstructured, tetrahedral elements[11,12]. Further refinement involved adding three prism layers with 0.1 mm thickness with a height ratio of 1 at the nasal airway walls.

A.

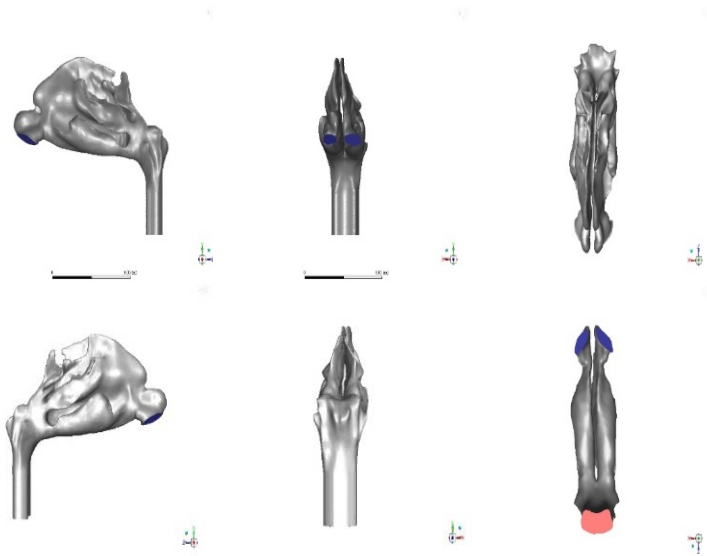

B.

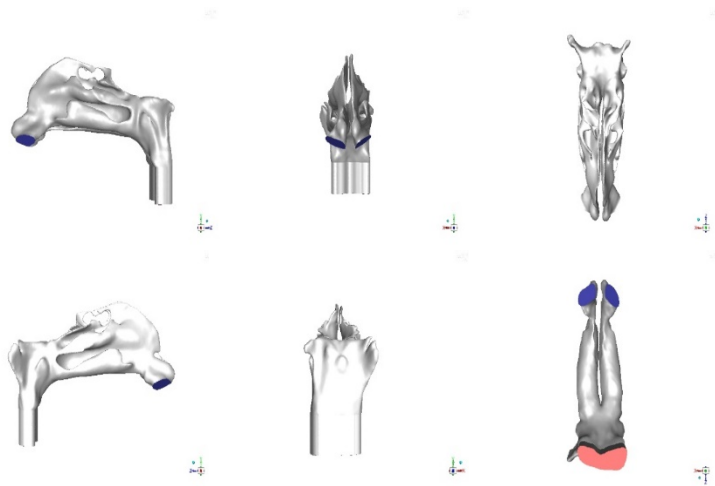

Figure 1. Panel A: Image of nasal cavity computational model from subject 1 (UNC003_Basu); Panel B: Image of nasal cavity computational model from subject 2 (UNC071_Basu)

**Inspiratory Airflow Simulation**

After digitization and meshing, the nasal geometry models were imported into Ansys FLUENT for inspiratory airflow simulations. We simulated two different airflow rates: 15 LPM (L/min) and 30 LPM. At the resting breathing rate, 15 LPM, a viscous laminar model was used to simulate airflow[13]. This scheme assumes that the fluid flows in smooth paths, with no swirling. The laminar model is not a good approximation for higher flow rates in the nasal cavity, due to effects of turbulence caused by the geometry of the airway. Since higher flow rates in the nasal cavity are better modeled by taking turbulence into account, the Large Eddy Simulation (LES) scheme was used to achieve the target output airflow rate of 30 LPM. LES often cannot resolve behavior at small length scales; instead, sub-grid scale models are used to compensate for these fluctuations. The Kinetic Energy Transport model was used as the LES sub-grid scale model[14].

The inspiratory airflow simulation was based on the principles of conservation of mass and conservation of momentum under the assumptions of steady state conditions and constant airflow density (incompressible flow). ANSYS Fluent numerically solves these equations for the chosen nasal geometry model, taking into account the specified laminar or turbulence scheme. The solver uses a SIMPLEC scheme, least squares cell based gradient, bounded central differencing momentum, and second order upwind for spatial discretization.

The flow is pressure driven, with the pressure at the outlet (the nasopharyngeal outlet plane) set to -50 Pa compared to atmospheric pressure at the inlets (the nostrils). The inlets were set to reflect particles back through the airway, simulating the effects of inhalation. The surfaces of the nasal cavity are categorized as walls with zero velocity, causing the air near these walls to similarly have zero velocity (no slip condition). This condition also implies that particles can get trapped at the walls.

Air density and viscosity were set as 1.204 kg/m^3 and 1.825e-05 kg/m-s respectively. Based on existing protocol, the time step for the 30 LPM LES simulation was 0.0002 second[15]. The final result is a velocity or pressure field at every point in the airflow; the solution converges based on residuals for mass continuity and velocity. Additionally, the mass flow rate and pressure at the outlet were monitored for stability.

**Drug Particle Tracking Simulation**

The airflow simulation for each nasal cavity model was evaluated in terms of its receptiveness to the simulated drug release. This trace of the drug particle trajectories confirmed whether or not the drug reached the nasopharynx, the first site of infection. The trajectories of the drug particles were tracked under the conditions of the aforementioned ambient airflow and followed the "discrete second phase model in a Lagrangian frame of reference" [16]. Using the modeling capabilities of FLUENT, spherical particles depict the nasal spray droplets that are disseminated in continuous airflow phase[17] as modeled by the particle force balance equation. This equation as shown below captures the behavior of a single particle under the equalization of particle inertia and its acting forces and is later numerically integrated to account for the aggregate of drug particles released[16,17].

$$\frac{du_d}{dt} = \frac{18\mu}{d^2 \rho_d} \frac{C_D Re}{24} (u - u_p) + g\left(1 - \frac{\rho}{\rho_d}\right) + F_B \tag{1}$$

The parameters that comprise this equation includes the airflow phase velocity, particle velocity, gravitational acceleration, all other relevant body forces (such as Saffman lift force which will push the particles transverse to the airflow direction as attributable to shear force) applied on the particle, densities of the particle, and air as indicated by the respective terms/variables: u, $u_d$, g, $F_B$, $\rho_d$, $\rho$ [16,17]. For further clarity, the relation between the particle inertia and its acting forces is computed as the sum of the drag force per unit particle mass (where $C_d$ is the drag force coefficient, d is the diameter of the particle, and Re is the Reynolds number), force of gravity on the particle, and any other additional forces applied on the particle.

For the purposes of this study, the drug particles were simulated and tracked using two different types of injections: surface injection and solid cone injection. It was incumbent on the simulations to best replicate the deposition of droplets in various pattern configurations. Surface injection emulates a release of droplets from each selected surface, consequently creating a particle stream of droplets[17]. Comparably, the solid cone injection would mimic the behavior of a nasal spray injection with particles being released from a single point and ejecting in a hollow-cone spray type fashion. To maintain uniformity within simulated model properties, the particle velocity and the total flow rate was set to 10 m/s and 1 x 10$^{-20}$ kg/s for all simulations respectively.

The Valois VP7, a commercial nasal spray pump currently in the market was used as the basis for all simulations. The device is specifically designed for the use of ethical drug products so consequently its performance was analyzed on whether it would be conducive to the needs of the project in terms of drug release [18]. The plume angle of the nasal spray ranged from 20°- 60° and the angle chosen was contingent on the droplet size. The plume angle utilized was 55.86° (half-cone angle of 27.93° in Fluent). Under these parameters, the particle tracking of two different cone positions (current use and perturbed direction) in addition to different particle sizes was simulated. All particle tracking was set up under laminar conditions with a flow rate of 15 LPM and turbulent airflow conditions with a flow rate of 30 LPM in its respective simulations.

**Surface Injection**

Surface injection was one of the types of injections used to evaluate the release of drug particles from selected surfaces. This injection use added a level of control as the number of particles injected was dependent on the number of points on the planes. As a means of standardization within computations nostril (inlets) planes were selected as the injection surface and particle sizes with uniform distribution of 1-24 μm were employed. The particle type selected was inert (with its medium of material being anthracite) as that choice accounted for a discrete phase element that followed the conventions of the force balance equation.

Following the numerical properties as prepared by Saikat Basu in the protocol for inert particle - DPM-based tracking, the tracking element was set to a maximum number of steps equal to $1 \times 10^6$ and a step length factor of 20. The numerical features of the discrete phase model incorporated a tolerance of $1 \times 10^{-8}$, maximum number of refinements of 20 and a tracking scheme of a higher order state/scheme as Runge-Kutta and a lower order state/scheme as implicit. The boundary conditions acted in accordance with those described earlier as a stationary wall (reflect type for nostrils; trap type for walls; escape type for outlet), and no slip for shear condition. The x-vel, y-vel, z-vel was set as 0 m/s across all terms and the simulation was performed with a density of 1500 kg/m3.

**Solid Cone Injection**

The solid cone injections encompassed the release of particles from a single point and as a result mirrored a nasal spray like spray deposition. To assess the robustness of particle release, different spay directions were used and their corresponding spray deposition were tracked and analyzed. Two simulations were conducted to demonstrate the extent to which the nasal spray can release the drug particles. The first simulation involved tracking particles undergoing two different cone positions with uniform distributed particle size from 1-24 μm similar to the procedure done for surface injection.

The two cone positions implemented was the current use (CU) and the perturbed direction (PD). These injection point positions were identified by first locating the center point of the nostril plane and then applying the insertion depth and angle of the nasal spray so that the new coordinate values for the injection point can be calculated (under the assumption of changes only in the y and z direction). The measurement of the positional injection point is exemplified as shown in figure 2.

For further explanation, the injection point and the center point of the nostril plane matched in x-coordinate value as per the premises of no change in the x-direction. The change in the y-direction was set as the product of the known insert depth and the sine of the insert angle. The designated insert depth and angle were 5 mm and 22.5° for the purpose of this study. In a consistent manner, the change in the z-direction was set as the product of the insert depth and the cosine of the insert angle. This resulted in a newly established point of reference for the injection point, center point on the nostril plane, as well as its change in direction. The two positions only differed in their cone direction. The cone of CU positions aligned directly to the direction of the insert angle (22.5°) and had lacked a vertical angle. Conversely, the PD position was set toward the direction of the airway with the intention of improving delivery efficiency. This difference in orientation is demonstrated in figures 3-6.

The second simulation involved tracking particles undergoing two different cone positions with the varying range of particle sizing. This measurement was run with the Rosin-Rammler diameter distribution of the following ranges of particle size: 1 - 10 µm, 10-20 µm, 20-30 µm, 1-20 µm, 1-30 µm, 1-80 µm, 10-90 µm, 20-100 µm, 30-110 µm.

Following the numerical properties as prepared by Saikat Basu in the protocol for inert particle - DPM-based tracking, the tracking element was set to a maximum number of steps equal to $1 \times 10^6$ and a step length factor of 20. As reflectant in the conditions of the particle force balance equation, the physical model permitted the Saffman Lift Force term accordingly. The numerical features of the discrete phase model incorporated a tolerance of $1 \times 10^{-8}$, maximum number of refinements of 20 and a tracking scheme of a higher order state/scheme as Runge-Kutta and a lower order state/scheme as implicit.

The particle type selected was inert (with its medium of material being anthracite) as that choice accounted for a discrete phase element that followed the conventions of the force balance equation. In addition to this, the variable for the x-position, y-position, and z-position was set in relation to the model being used for both the CU and PD cone position (these values would differ between positions as desired). Like the surface injection, this simulation was performed with a density of 1500 kg/m3.

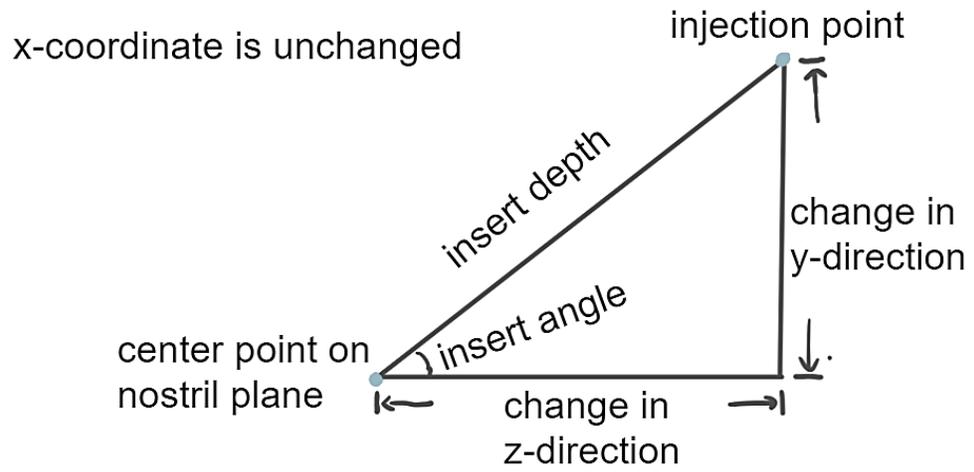

*Figure 2.* Calculation of the position of the injection point

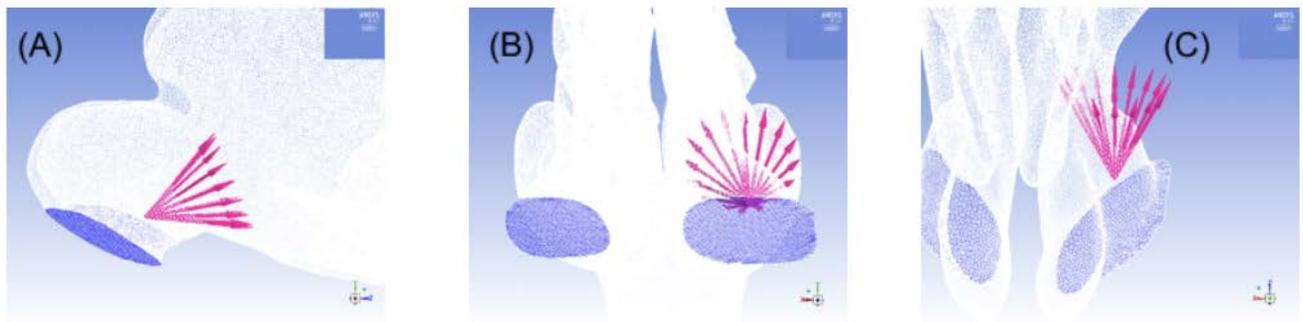

Figure 3. Cone Injection of Current Use Position Setup (A: Side, B: Front, C: Top) for Nasal Geometric Model UNC003_Basu

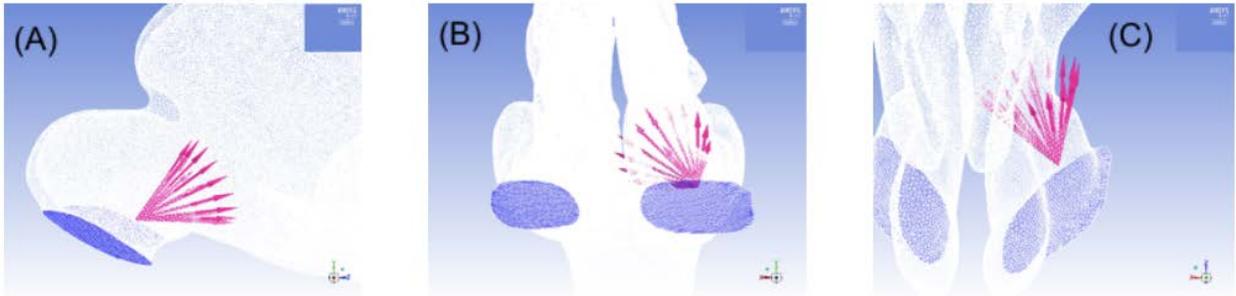

Figure 4. Cone Injection of Perturbed Direction Position Setup (A: Side, B: Front, C: Top) for Nasal Geometric Model UNC003_Basu

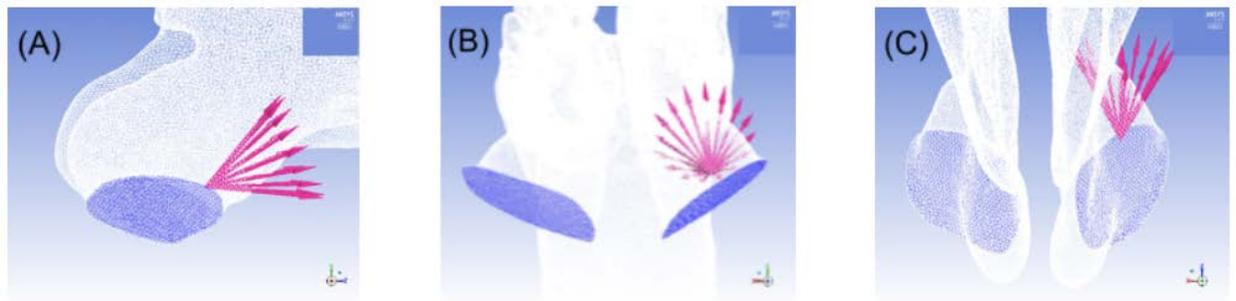

Figure 5. Cone Injection of Current Use Position Setup (A: Side, B: Front, C: Top) for Nasal Geometric Model UNC071_Basu

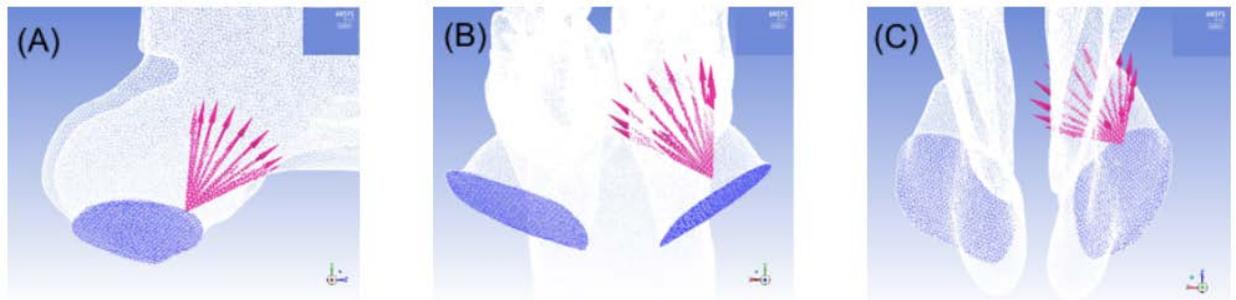

Figure 6. Cone Injection of Perturbed Direction Position Setup (A: Side, B: Front, C: Top) for Nasal Geometric Model UNC071_Basu

## 3. Results

**Surface Injection**

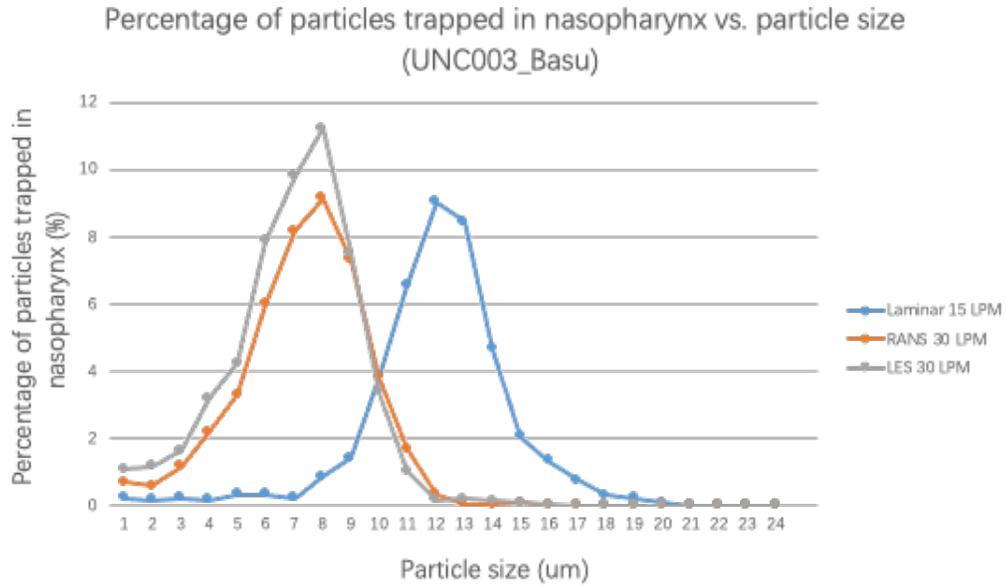

Figure 7: Graphical representation of the percentage of particles trapped in the nasopharynx vs. particle size for nasal cavity UNC003_Basu done using Laminar 15 LPM and LES 30 LPM flow rate. The distribution for laminar 15 LPM simulation is observed to have a bell curve shape. The distribution for LES 30 LPM simulation is skewed to the right.

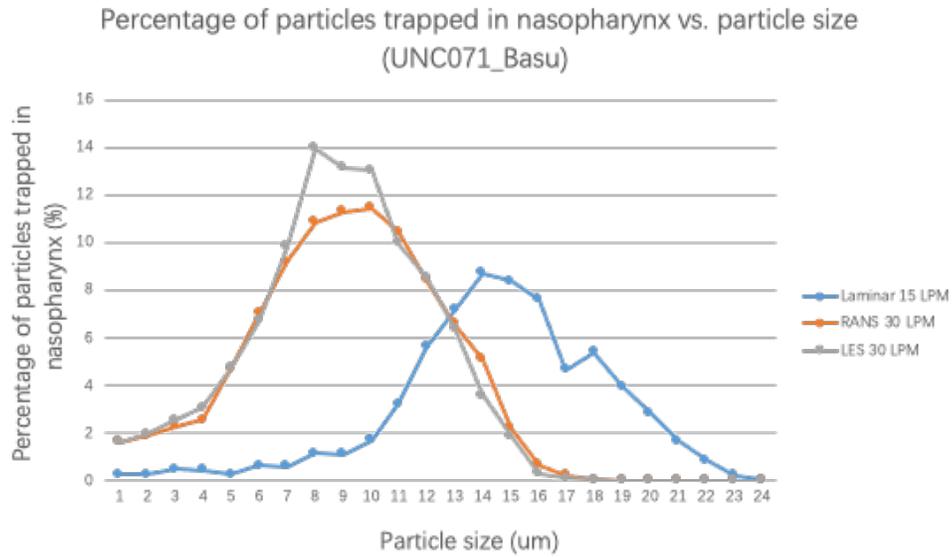

Figure 8: Graphical representation of the percentage of particles trapped in the nasopharynx vs. particle size for nasal cavity UNC071_Basu done using Laminar 15 LPM and LES 30 LPM flow rate. The distribution for laminar 15 LPM simulation is observed to have a bell curve shape. The distribution for LES 30 LPM simulation is skewed to the right.

**Cone Injection with Ranged Size**

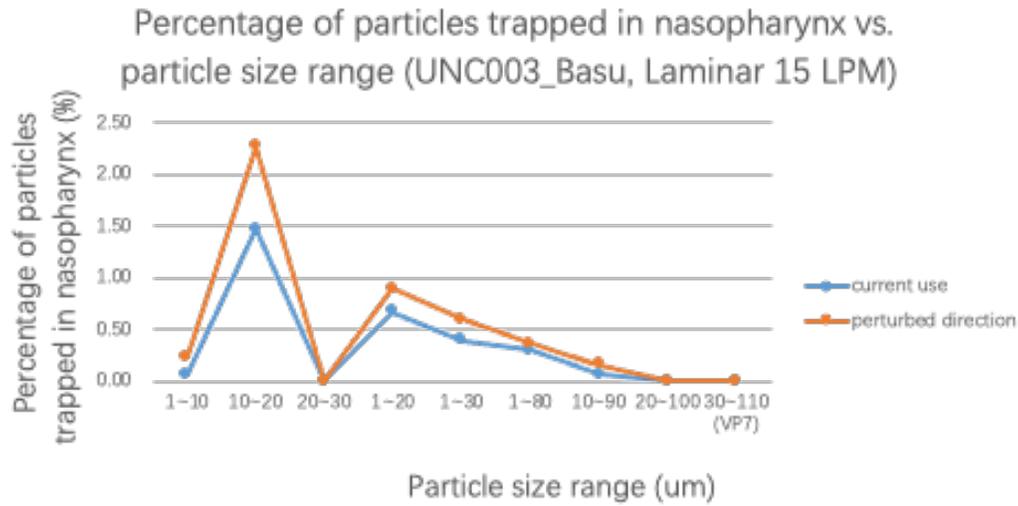

Figure 9: Graphical representation of the percentage of particles trapped in the nasopharynx vs. particle size for nasal cavity UNC003_Basu done using Laminar 15 LPM flow rate. It's observed that with the parameters of VP7 (the current commercial nasal spray product on the market), no particle can be delivered to nasopharynx. The maximum percentage of particles trapped in the nasopharynx occurs at 10-20um. There is no major difference observed between the current cone injection method and the perturbed direction method.

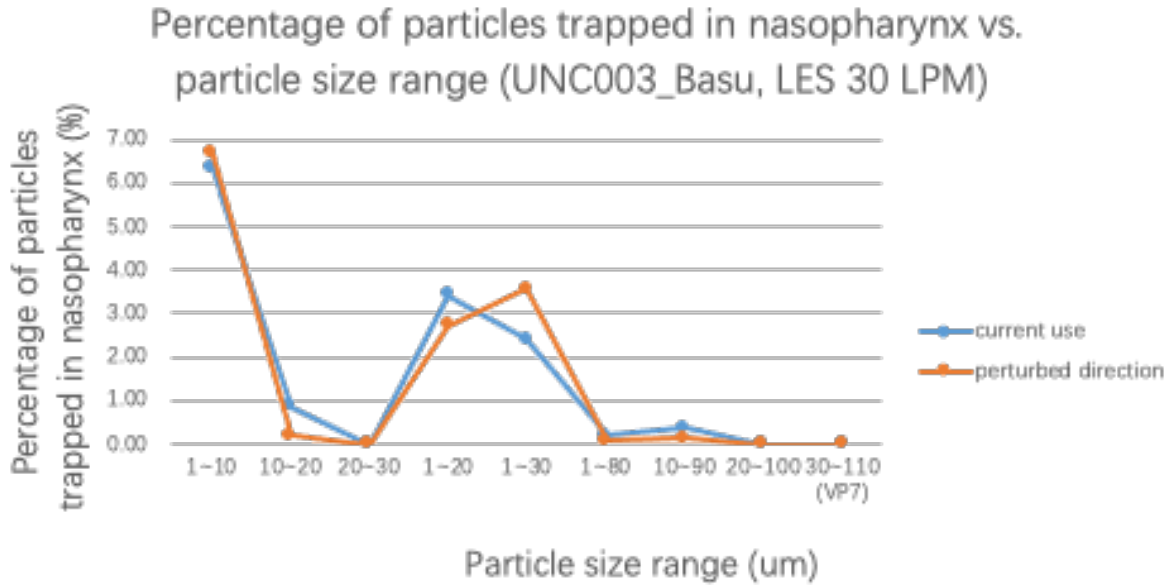

Figure 10: Graphical representation of the percentage of particles trapped in the nasopharynx vs. particle size for nasal cavity UNC003_Basu done using LES 30 LPM flow rate. It's observed that with the parameters of VP7 (the current commercial nasal spray product on the market), no particle can be delivered to nasopharynx. The maximum percentage of particles trapped in the nasopharynx occurs at 1-10um. There is no major difference observed between the current cone injection method and the perturbed direction method.

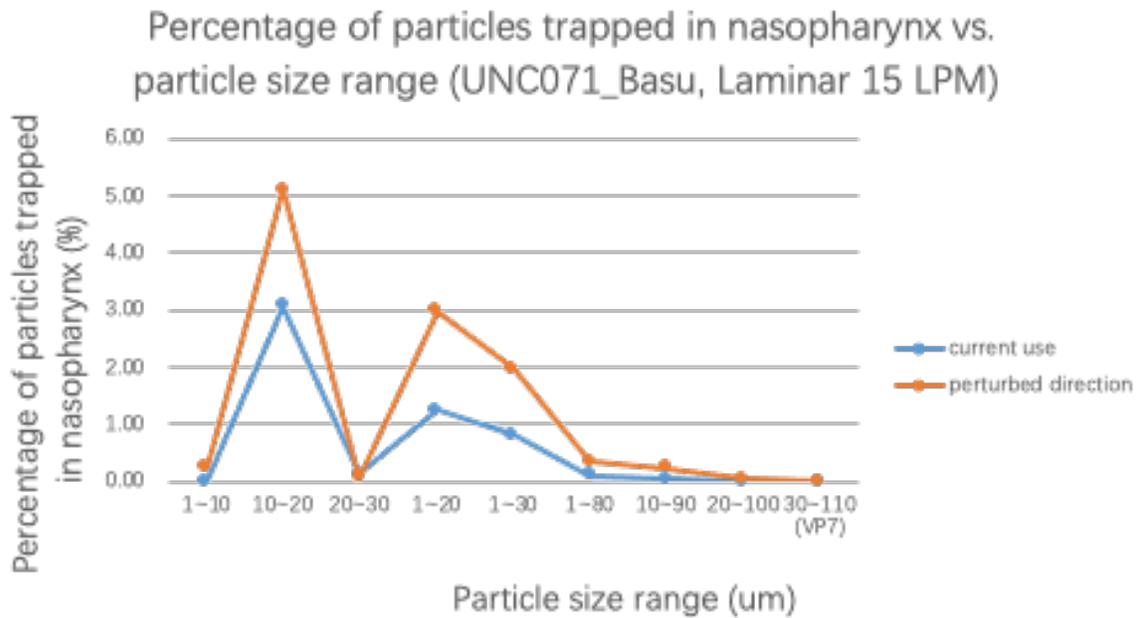

Figure 11: Graphical representation of the percentage of particles trapped in the nasopharynx vs. particle size for nasal cavity UNC071_Basu done using Laminar 15 LPM flow rate. It's observed that with the parameters of VP7 (the current commercial nasal spray product on the market), no particle can be delivered to nasopharynx. The maximum percentage of particles trapped in the nasopharynx occurs at 10-20um. There is no major difference observed between the current cone injection method and the perturbed direction method.

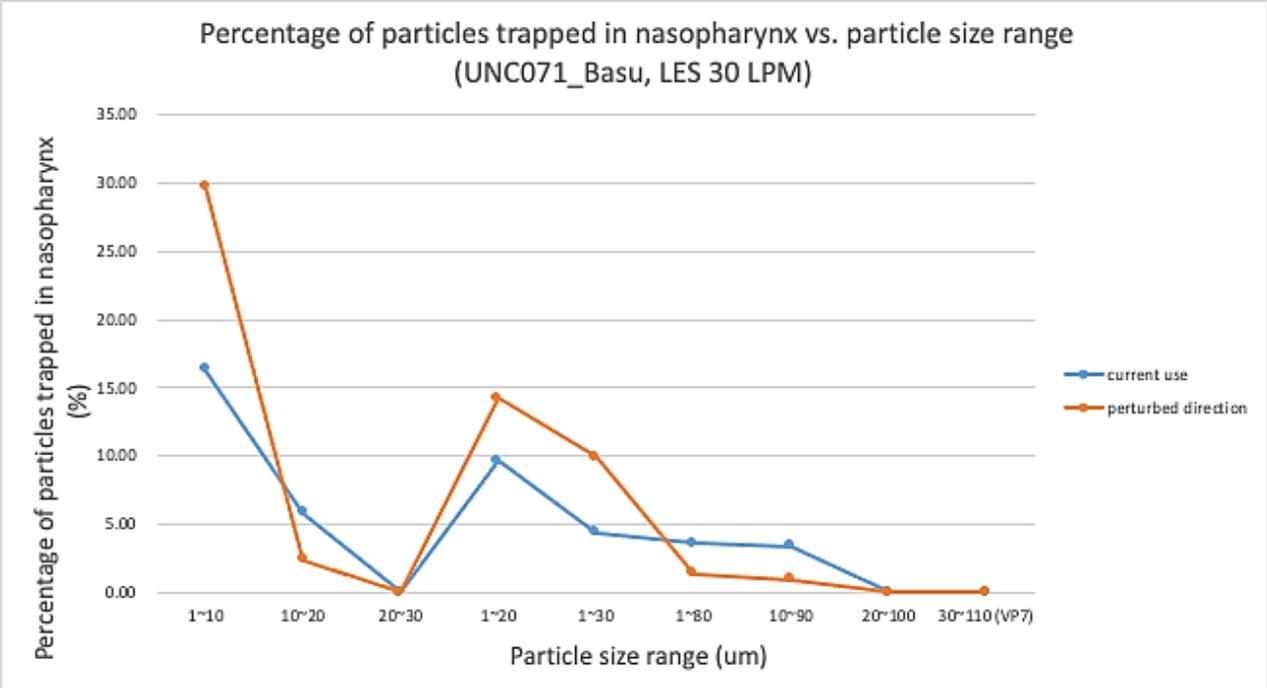

Figure 12: Graphical representation of the percentage of particles trapped in the nasopharynx vs. particle size for nasal cavity UNC071_Basu done using LES 30 LPM flow rate. It's observed that with the parameters of VP7 (the current commercial nasal spray product on the market), no particle can be delivered to nasopharynx. The maximum percentage of particles trapped in the nasopharynx occurs at 1-10um. There is no major difference observed between the current cone injection method and the perturbed direction method.

**Cone Injection with Particle Size Distribution**

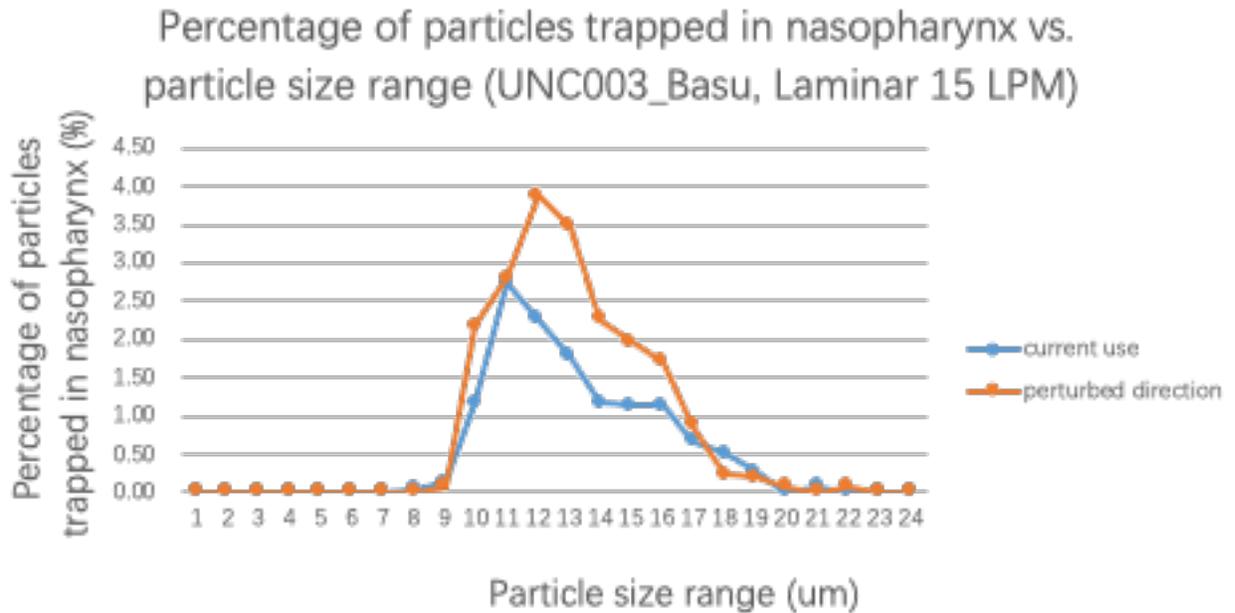

Figure 13: Graphical representation of the percentage of particles trapped in the nasopharynx vs. particle size for nasal cavity UNC003_Basu done using Laminar 15 LPM flow rate. The distribution is similar to that of the surface injection. There is no major difference observed between the current cone injection method and the perturbed direction method.

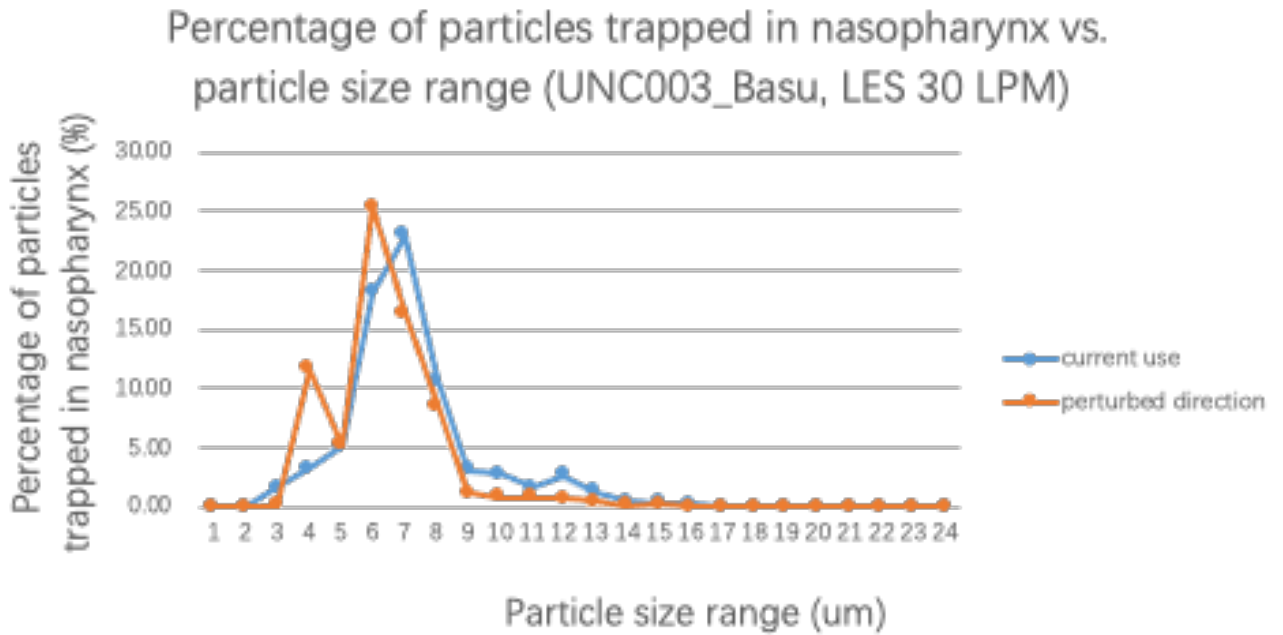

Figure 14: Graphical representation of the percentage of particles trapped in the nasopharynx vs. particle size for nasal cavity UNC003_Basu done using LES 30 LPM flow rate. The distribution is similar to that of the surface injection. There is no major difference observed between the current cone injection method and the perturbed direction method.

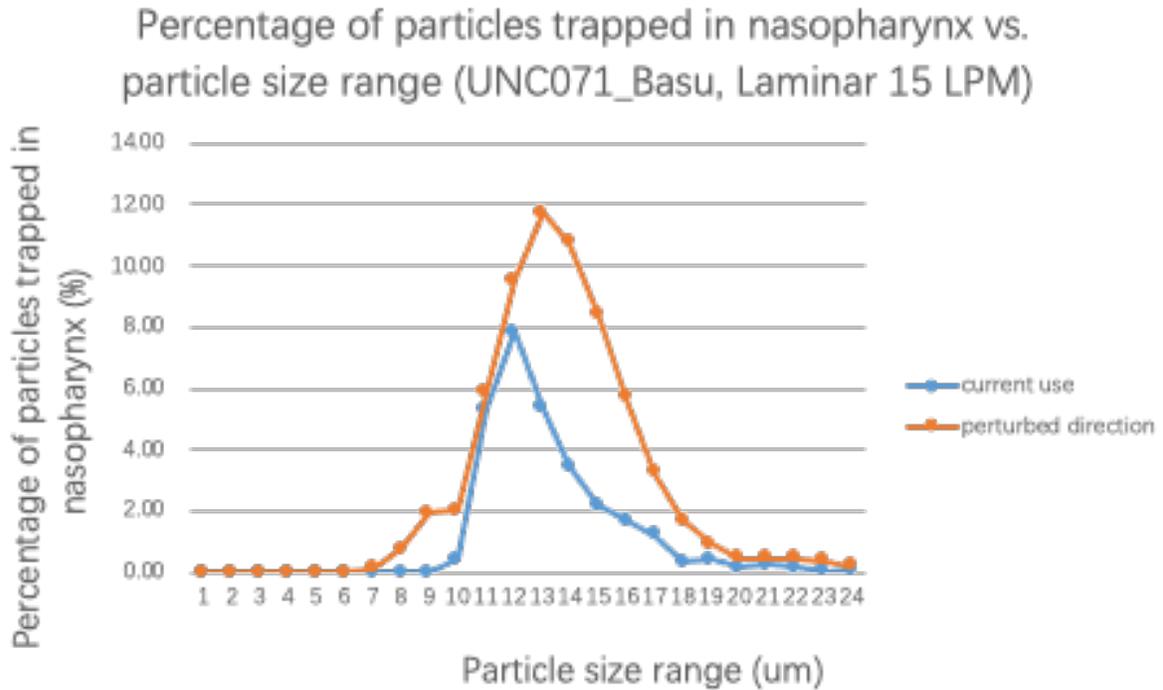

Figure 15: Graphical representation of the percentage of particles trapped in the nasopharynx vs. particle size for nasal cavity UNC071_Basu done using Laminar 15 LPM flow rate. The distribution is similar to that of the surface injection. There is no major difference observed between the current cone injection method and the perturbed direction method.

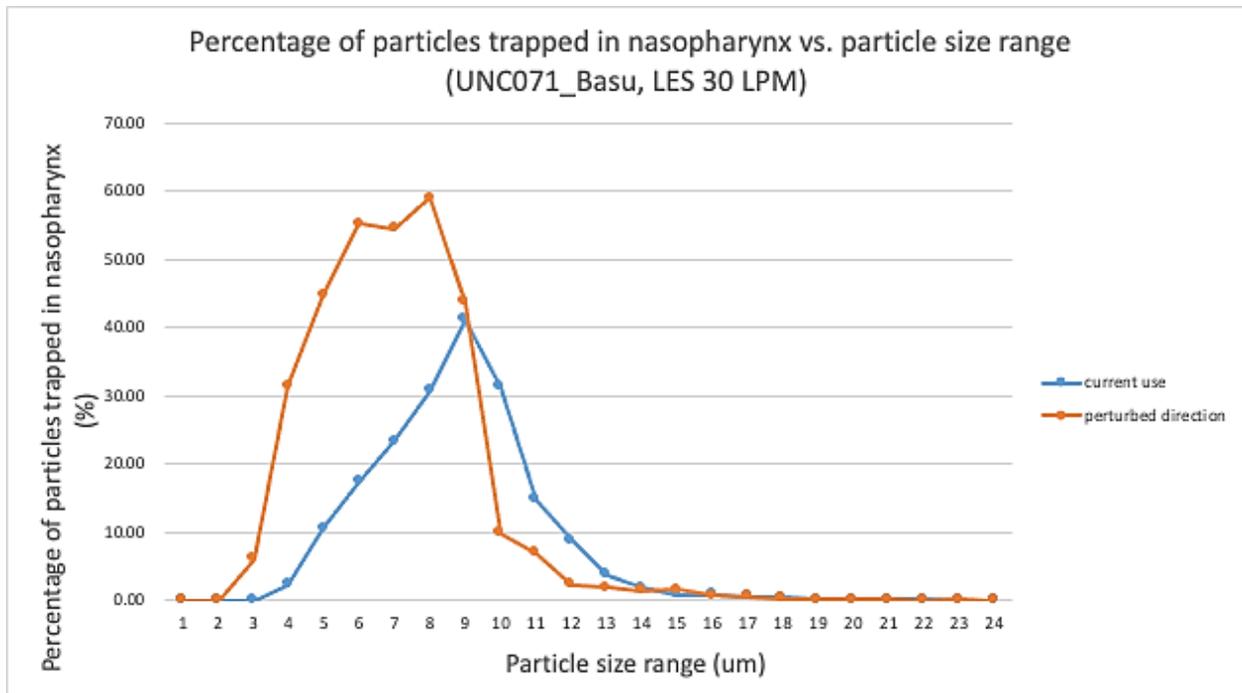

Figure 16: Graphical representation of the percentage of particles trapped in the nasopharynx vs. particle size for nasal cavity UNC071_Basu done using LES 30 LPM flow rate. The distribution is similar to that of the surface injection. There is no major difference observed between the current cone injection method and the perturbed direction method.

## 4. Discussion

The goal of this project was to begin developing an *in silico* prototyping tool that would allow identification of optimal parameters for nasal delivery of drug combinations targeting COVID-19. It was deduced from the results that for the surface injection model, the distribution of the percentage of particles trapped in the nasopharynx in relation to the particle size (for the laminar airflow condition with flow rate of 15 LPM) resembled a pattern close to a bell shape as depicted in figures 7 and 8. This implies that there is a normal distribution of percent particles being trapped in the nasopharynx amongst the varying particle sizes and that the data is symmetrical. The curve, thereby, allows for reasonable expectations towards the prospect of using different particle sizes and its corresponding spray reach in the nasopharynx.

As presented in figures 7 and 8, at large the distribution for simulations (LES) at 30 LPM flow are skewed right. This pattern was shared amongst both models (UNC003_Basu & UNC071_Basu). When measuring the differences between the particle size at which the maximum percentage of particles successfully trapped in the nasopharynx, it was determined that it occurred at a smaller particle size from the higher flow rate of 30 LPM (LES) than the lower flow rate of 15 LPM (laminar). Specifically, as exemplified in the data representing model UNC003_Basu in figure 7, the maximum percentage occurred at 12 µm, and 8 µm for the simulations under conditions of Laminar 15 LPM and LES 30 LPM respectively. Whereas for the data representing model UNC071_Basu in figure 8, the maximum percentage occurred at 14 µm and 8 µm for the simulations under conditions of Laminar LPM and LES 30 LPM. This range of particle size conveys the impression that optimal particle sizes for the most successfully displaced droplets would be around the values described earlier. Now, with all different simulations considered, the LES simulation under the conditions of 30 LPM flow rate held the largest percentages of particles (11.24%) trapped in the nasopharynx for model UNC003_Basu. This information is to be acknowledged when optimizing the drug release approach in the nasal spray.

The simulations falling under the cone injections with ranged sizing had no particles delivered to the nasopharynx regardless of cone position. This lack of spray deposition adds uncertainty to the effectiveness of the Valois VP7, the nasal spray pump currently in the market. As for the distribution of the data, it was found that generally a smaller range of particle size would elicit a higher percentage of particles successfully reaching the nasopharynx as reflected in figures 9 - 12. It can also be noted that the modified cone direction (perturbed) was not significantly superior to the current use cone direction, thus implying that the disparate cone positions were not substantially effective in increasing the number of particles trapped in the nasopharynx. This outcome persisted in the analysis of the cone injections with distributed sizing as well as shown in figures 13-16.

It was also identified for the cone injection with ranged sizing that the maximum percentage of particles successfully trapped in the nasopharynx occurred at a lower range scale (1-10 µm) for the larger flow rate of 30 LPM (LES) than the lower flow rate of 15 LPM (laminar) which occurred at the size range of 10-20 µm. The trend presented itself once again in the cone injection simulation with distributed sizing as shown in figure 13-16. This behavior matches our results from the surface injection simulation, thus indicating that the optimal particle sizes should be within 6-14µm (amongst all simulation types) in order to attain relatively large particle deposition.

Despite the consistent distribution patterns amongst the simulations, it is still incumbent to improve the results in order to significantly enhance the topical delivery release within the nasopharynx. One limitation of the simulation was centered around the nasal spray device in that the Valois VP7 was inapt at meeting the study desired results. To have a more effective delivery of drugs, it is imperative to have a nasal spray device that can produce droplets of smaller sizing. To account for this inefficiency, future improvements aimed at increasing the maximum percentage of particles trapped in the nasopharynx can be implemented. In conjunction with the CFD-based simulations, the numerical modeling of release parameters can be enhanced to ameliorate the performance of the drug release. Additionally, a more advanced method needs to be developed to determine the perturbed direction of the cone position. These future improvements of the delivery release and device alongside with additional testing on different models would promote efficiency in deposition patterns and optimal spray release points.